# Robust Topological Conduction in Bi$_2$–Bi$_2$Se$_3$ Superlattices at Ambient Conditions

Lakshan Don Manuwelge Don[1], Md. Sakauat Hasan Sakib[1], Gracie Pillow[1], Sara McGinnis[1], Seth Shields[2], Joseph P. Corbett[1]
1. Miami University, Department of Physics, Oxford, OH 45056
2. NSF NeXUS Facility, The Ohio State University, Columbus, OH, 43210

**Abstract**

Topologically protected surface states have garnered significant attention due to their robustness against perturbations and potential applications in optoelectronics. Bi$_2$-Bi$_2$Se$_3$ is a topological semimetal composed of a 2D bismuthene sheet and a Bi$_2$Se$_3$ quintuple layer, forming an intrinsic superlattice. This study investigates the electronic structure and edge states of Bi$_2$-Bi$_2$Se$_3$ [001] oriented films under ambient conditions through conducting atomic force microscopy (C-AFM). Point I-V spectroscopy and current imaging are used to characterize the surface and local transport properties of bismuthene and Bi$_2$Se$_3$ terminated layers. Our measurements reveal force-dependent shifts in conduction mechanisms in both bismuthene and Bi$_2$Se$_3$, transitioning from direct tunneling (DT) at low forces and low biases, to Fowler–Nordheim tunneling (FNT) at low forces and high biases, and eventually to a more ohmic-like behavior at the highest forces. Under DT conditions on the bismuthene termination, we observed the Dirac cone in the dI/dV spectroscopy. Edge states are observed along the perimeter of the (001) terraces for both terminations, and are observed to have higher conductivity than the local terrace. Force-dependent imaging revealed an increase in the width of the edge state as force increased, until the conductive edge state appeared to cover the entire terrace. Furthermore, terrace heights display a force dependent distortion from the high tip forces, which indicates that the transition to the ohmic-like contact regime on either termination results from a complex interplay between strain and tip induced effects. All measurements were performed under ambient conditions, which demonstrates the robustness of the topological and edge states to ambient conditions.

**Main:**

Topological quantum materials have unlocked entirely new paradigms in condensed matter physics, where electronic states are protected not by symmetry alone, but by global topological invariants [1–6]. These materials, including topological insulators, Weyl semimetals, and Dirac semimetals, host robust boundary states with spin-momentum locked electrons, highly spin-polarized surface currents, giant non-suturing linear magnetoresistance, and novel optoelectronic responses to circular polarized light. As the field of topological condensed matter matures, a central challenge is to move beyond idealized crystal models and probe how these topological states behave in complex environments and at device-relevant interfaces.

Intrinsic and engineered superlattice systems, such as the Bi$_2$–Bi$_2$Se$_3$ system investigated here, represent an emerging class of topological materials that allow unprecedented control over dimensionality and interlayer coupling [7–9]. Composed of alternating 2-dimensional (2D) bismuthene (Bi$_2$) layers and a quintuple layer (QL) of bismuth selenide (Bi$_2$Se$_3$) separated by van der Waals gaps (vdW), Bi$_2$–Bi$_2$Se$_3$ exhibits distinct surface terminations with markedly different topological surface states depending on the termination, see *Figure 1*. Two distinct unreconstructed surface terminations are possible within this material, bismuthene and Bi$_2$Se$_3$, each hosting a topological surface state with different Kramers point energies and dispersions [7]. These differences arise from interlayer interactions between the bismuthene and Bi$_2$Se$_3$ layers. The bismuthene is a 2D topological semimetal that supports a Dirac cone centered at the Fermi level, whereas the QL slab of Bi$_2$Se$_3$ hosts a topological surface state within the valence band but with a parabolic-like dispersion. This unique feature of Bi$_2$-Bi$_2$Se$_3$ offers a platform to study distinct topological surface states within the same material [7,8]. This internal heterogeneity makes Bi$_2$–Bi$_2$Se$_3$ an ideal platform to investigate how topology manifests at atomic interfaces and edges.

Research on Bi$_2$-Bi$_2$Se$_3$ superlattices spans growth mechanisms, electronic structure, and functional properties. DC magnetron sputtering under external magnetic fields has enabled tunable film morphology, producing structures ranging from smooth layered films to faceted granular morphologies as a function of sputtering power and pressure, producing large single crystal domains spanning 100's of microns in size [8]. Nucleation studies of DC sputter growths revealed an initial Volmer–Weber growth mode—characterized by isolated nanoplatelet formation—followed by a pressure-dependent transition to Frank–van der Merwe growth. Radial distribution analysis from AFM data showed nearest-neighbor ordering during nucleation enabling diffusion length estimates [10]. Molecular beam epitaxy (MBE) studies precisely controlled the beam flux Bi:Se ratio and growth temperature, enabling growth modes from van der Waals' condensation to spiral growth, thus achieving single-crystalline Bi$_4$Se$_3$ of dislocation-free microplate or mounded thin-film morphologies [20]. Such tunability over growth parameters, and the resulting control of film morphology, forms an essential bridge between fundamental materials studies and the realization of high-performance devices.

Band structure investigations using angle-resolved photoemission spectroscopy (ARPES), spin-resolved ARPES, and photoelectron emission microscopy (PEEM) have shown that the Bi-terminated surface hosts nearly linear Dirac states near the Fermi level, whereas the Bi$_2$Se$_3$-terminated surface exhibits parabolic dispersing states in the valence band. These termination-dependent features were validated both experimentally, using termination selection through photoelectron emission microscopy, and theoretically by comparison to spin-dependent DFT simulations, confirming the topological nature of the observed states [7,21]. STM investigations confirmed the presence of topologically protected surface states in sputtered films through scanning tunneling spectroscopy measurements of the two terminations, demonstrating compatibility with industrially scalable growth approaches [8]. UV-Vis spectroscopy detected bandgap expansion with reduced nanoplatelet size, and observing a transition from 2D confinement to full 3D confinement in line with DFT models [10].

From a device standpoint, Bi$_2$-Bi$_2$Se$_3$ has demonstrated favorable interface behavior, forming Ohmic contacts with n-type and intrinsic Si(111), and Schottky contacts with p-type Si(111), enabling efficient silicon-based photodetectors [9]. The material's ultralow lattice thermal conductivity (0.24–0.27 W/mK), even lower than that of $Bi_2Se_3$ (~0.35 W/mK), is attributed to enhanced phonon scattering across van der Waals (vdW) interfaces, while retaining good carrier mobility—properties that make it attractive for thermoelectric applications [22]. Furthermore, the system exhibits complex interlayer bonding, with stronger non-covalent, vdW interactions between bismuthene bilayers and quintuple layers (8.6 kcal/mol) than between adjacent quintuple layers (6.3 kcal/mol). The relatively low sliding energetics (1–2 kcal/mol) suggests design flexibility in Bi–Se-based materials for future spintronic and energy-harvesting devices [23]. Kinetically controlled MBE films reported semimetallic bipolar conduction, contrasting with the n-type semiconducting behavior of $Bi_{(2+x)}Se_3$ ($1.7 < x < 2$). These differences arise from stacking disorder, which alters the material's electronic and thermoelectric behavior [20].

Yet a key question remains: how resilient are the bismuthene and bismuth selenide terminations, and their topological states under realistic device conditions, where local strain, variable contact forces, and ambient oxidating environments come into play? To date, local transport measurements that directly probe such effects have been scarce.

Here, we combine conductive atomic force microscopy (C-AFM) with current-voltage (IV) spectroscopic analysis to map the spatially resolved transport characteristics and topological states of $Bi_2$–$Bi_2Se_3$ under ambient conditions. We reveal force-dependent transitions in the tip-sample conduction mechanism with termination-specific electrical properties, and observe robust localized edge states at the terrace boundaries. Our findings highlight the opportunities and challenges in unraveling local electronic details for harnessing topological surface states for scalable device applications.

These findings provide valuable insights into the spatially resolved electronic transport properties of intrinsic $Bi_2$-$Bi_2Se_3$ superlattice under ambient conditions which demonstrate robustness to atmospheric conditions without the loss of their topological nature and contribute to a broader understanding of topological materials interfaces under varying electrical contact conditions.

**Methods**
   A. **Sample Growth:**

In this study, films were grown using a custom-built ultra-high vacuum (UHV) coil-assisted sputtering epitaxy system. Samples are introduced into the system via a commercial transferable heater stage from Bricada Inc. through a load lock chamber with a $1 \times 10^{-8}$ Torr base pressure and transferred into the UHV growth chamber with a $1 \times 10^{-10}$ Torr base pressure. The substrate is placed in the center of the deposition chamber, between two 1.3-inch FerroTec MAK magnetron guns. The transferable heater stage is connected to a 4-axis goniometer allowing for three translation motions (X,Y,Z) and rotational motion (Θ) to face either magnetron gun. Each magnetron is surrounded by an external set of Helmholtz coils in coaxial arrangement. A uniform magnetic field is generated between the substrate and the target through the coils to modify and confine the plasma during sample growth.

Sputtering targets were produced in-house 1.3-inch $Bi_2Se_3$ target using commercial powders of 99.999% pure Bi and 99.9% pure Se were from Elemental Metals. A manual hydraulic laboratory press was used to press powdered raw materials under high pressure of 40 MPa. The target subsequently sintered under vacuum ($3 \times 10^{-7}$ Torr) was performed at 100°C in a tube furnace.

The $Bi_2$–$Bi_2Se_3$ films were deposited on basal-plane-oriented sapphire substrates from MTI Corp. Prior to growth, the sapphire substrate was heated to a growth temperature of about 250°C for 40 minutes to stabilize it. Temperature calibration was performed by recording the heater surface temperature as a function of applied heater power using a thermocouple. Ultra-pure Argon gas was introduced into the system via a UHV leak valve. Growths were performed at an Ar pressure of 5 mTorr. A DC power of 5 W is supplied to the $Bi_2Se_3$ target with a confining 15 mT magnetic field. Films were grown to a thickness of 500-900 nm. Samples produced for STM investigations were capped in ~100 nm Se cap were deposited to protect the sample during transit. Decapping occurred via sputtering and annealing.

### B. Characterization:

Atomic force microscopy imaging was performed in an AFMWorkshop commercial high-resolution multimodal instrument. Topographic imaging alongside simultaneous conducting mode imaging, and point-specific I-V measurements were collected utilizing contact mode with commercial conducting silicon probes with platinum overall coating (ElectriCont-G) from BudgetSensors. The system is wired as to have the bias is applied to the tip and the sample is at virtual ground. Imaging processing was performed using the open-source software Gwydion [24]. Topographic imaging of a region consisting of atomically smooth terraces allowed for step height analysis to be performed. Comparison of the measured and predicted step heights from the crystal model allowed for the determination of the surface termination. Local transport properties were measured by performing I-V sweeps at specific points on both surface terminations. dI/dV plots were produced numerically using a 5-point derivative method. To clarify the conduction mechanisms, the multiple plotting procedures (e.g. semi-log and log-inverse) were utilized on I-V spectroscopy data to ascertain the functional form and observe characteristic transitions in conduction mechanism. A qualitative analysis was performed by fitting the data to the direct tunneling (DT), Fowler-Nordheim tunneling (FNT), and Ohmic models [25–28]. The thermionic emission (TE) model was used to determine the local barrier height and ideality factor by considering the exponential behavior at high bias voltages.

STM measurements were performed at the NSF NeXUS facility on a Scienta Omicron INFINITY STM, equipped with a closed cycle cryocooler. All measurements were performed at ~14 K with an etched W tip. The tip was heated in UHV prior to measurements to remove the native oxide layer. Bias is applied to the sample. Prior to STM imaging, the samples were lightly $Ar^+$ sputtered (1 kV, normal incidence) and then heated to a nominal 140-150° C, as measured by a thermocouple placed near, but not on the sample. This procedure removed the Se capping layer from the samples. Differential conductance (dI/dV) maps were acquired simultaneously with topographic measurements using a standard lock-in technique.

**Discussion:**
**$Bi_2$-$Bi_2Se_3$ Terrace Characterization:**

Following our previous works, we prepared films that simultaneously contain atomically smooth layers oriented in the (001) direction alongside sparely populated nano-sized polyhedral crystal forms [8]. Figure *2*(a) shows a large scale image showing atomically smooth layers with some isolated crystal forms [10]. The smooth layers are represented by the deep purple contrast, where isolated crystal forms which shown pronounced heights up to ~300 nm are the bright yellow and orange contrast. Figure *2*(b) shows a zoomed-in region of an atomic staircase demonstrating the (001) stacking of the layers. Two linecuts are taken in Figure *2*(b), one across the staircase (blue line), and one along a staircase (red line). The linecuts in Figure *2*(b) are displayed in Figure *2*(c). Both possible terminations of $Bi_2$-$Bi_2Se_3$, the bismuthene and the $Bi_2Se_3$ layers, are observed. The bottom most terrace has a combination of both $Bi_2Se_3$ and bismuthene terminations, as shown by the red line showing a bismuthene step height of $0.21 \pm 0.1$ nm. Moving across the staircase, we can track the termination by measuring the step heights, where a $Bi_2Se_3$ step is $0.8 \pm 0.1$ nm, and a $Bi_2Se_3$+$Bi_2$ Step is $1.1 \pm 0.1$ nm, which enables a determination of the termination as performed in our previous works [7,8,10].

To corroborate microscopy imaging of the (001) layers, we performed a coupled symmetric XRD scan of the $Bi_2$-$Bi_2Se_3$ film, see Figure S *1* in the supplement. We observe a pronounced 001 family of peaks, as was the case with our previous work [8]. We targeted growth conditions such that the film would grow predominantly with atomically smooth 001-oriented crystal forms. We do observe weak higher index reflections, such as the (107) orientations, which are known to form as inclusions from our previous work [8].

We utilized force dependent imaging to assess impacts of tip induced strain and terrace height variations in imaging. Prior work on $WS_2$ has shown that under dynamic mode operation, the tip force influences the apparent height of monolayers, and this has been attributed to capillary forces from ambient water [29]. While we are using traditional contact mode methods in this work, sufficiently high loading forces can cause local strain in the film, which leads to distortions in the imaging, and potential difficulty in assigning termination due to measured atomic layer heights.

Figure *3* shows AFM images and linecuts under the various load conditions explored in this work. Figure *3*(a,b,c) are images selected at low (15 nN), mid (108 nN), and high forces (230 nN) of a terrace staircase containing both a terminations. Linecuts are marked in Figure *3*(a,b,c) as a colored lines (green, light green, and black), with additional linecuts taken under the full range of force conditions displayed in *Figure 3*(d). At low loading forces the imaging is crisp, with clearly defined terrace edges as seen in both the imaging and linecuts. As the force increases, small variations within the terrace height can be observed, but the more prominent effect of the increase loading is the loss of definition in the step edges. At the highest forces, the terrace structure is barely visible, as a significant distortion appears in the imaging, where the linecuts do not show clear steps, but rather a distorted slope. Interestingly, the transition to an Ohmic I-V characteristic, as discussed below, corresponds to the same applied forces where the terraces are no longer readily visible, indicating that local strain may play a role in the conduction regime of the tip-sample junction.

**$Bi_2$-$Bi_2Se_3$ Termination and Point IV-Spectroscopy:**

Beyond step height and crystal structure analysis, local point I-V spectroscopy is used to characterize the electronic structure of the bismuthene and $Bi_2Se_3$ layers. Our previous work using STM to perform point I-V spectroscopy [CITATION] identified a Dirac cone on the bismuthine termination, in good agreement with DFT and ARPES results [CITATION]. In an analog to our previous work, here C-AFM is used to measure the local I-V characteristic of the bismuthene and $Bi_2Se_3$ terminations. Figure *4*(a,d) shows another two staircases of $Bi_2$-$Bi_2Se_3$ layers with a corresponding across-staircase linecut (red line) to determine a bismuthene and termination from step height analysis. Similar to Figure *2* we measure the height across several bismuthene and $Bi_2Se_3$ steps to ascertain the local stacking order. Point-specific spectroscopy was taken as a function of applied force spanning an order of magnitude in force from 15 nN to 250 nN on the bismuthene and $Bi_2Se_3$ terminations.

Unlike STM, which operates exclusively in the direct tunneling region, C-AFM can probe a variety of conduction mechanisms including the direct tunneling regime, ohmic regime, thermionic emission, and FNT regime as a function of force and bias voltage [26,27,30]. Here we characterize I-V spectroscopy measurements to ascertain the conduction mechanisms by varying the force and voltage then assess the characteristic functional form of the I-V spectroscopy to make a mechanism determination.

These I-V curves are displayed in Figure *4*(c,f), alongside the location of spectroscopy indicated by the black X, in both the image and the corresponding linecut. The I-V curves are color coded to correspond to the applied force on both the bismuthene and $Bi_2Se_3$ terminations. The shape of the measured I-V curve changes dramatically as the force is changed by an order of magnitude.

The I-V spectroscopy has been plotted in a variety of formats in order to enable of assessment of the functional form of the data under various force and bias setpoints, which will allow for identification of the conduction mechanism. Figure *5*(a,d) shows the I-V measurements in Figure *4*(c,f) but plotted on a semi-log scale with an arbitrary vertical offset to display the data in a waterfall plot for bismuthene termination Figure *5*(a), and the $Bi_2Se_3$ termination, Figure *5*(c). For the bismuthene termination, *Figure 5*(a), there is shift in the Fermi level towards empty states as the applied force decreases. This dramatic shift in Fermi level is tentatively attribute to significant tip-induced band bending [31]. This is also observed in the plain I-V spectroscopy, *Figure 4*, but it is difficult to observe when plotted on a linear, as opposed to semi-log, scale. Similarly, the $Bi_2Se_3$ termination, *Figure 5*(d), also demonstrates a shift in the Fermi level as a function of applied force. At low forces the I-V spectroscopy is largely gapped, and as the force increase the apparent gap size decreases dramatically until completely pinching off at high loading forces. The center of the gap shifts around the zero until at the largest loading forces it become symmetric around zero bias.

For the bismuthene termination, we observe a transition in conduction mechanism as a function of force, which can be seen in our semi-log and log-log I-V spectroscopy in Figure 5(a-c). Beginning with the semi-log I-V plot, *Figure 5*(a), at the lowest forces, the data is asymmetric around zero bias. At the lowest forces the curves in the semi-log IV-spectroscopy data is asymmetric, with negative bias exhibiting a nonlinear response, this corresponds to electrons leaving the tip and entering states in the samples, which for the bismuthene termination are topological. Whereas at positive bias, the shape is nonlinear only at low biases, and then becomes linear after a ~0.15 V where electrons in the bismuthene leave the sample and enter trivial Pt metallic states. We tentatively attribute this asymmetry to potentially reflect the asymmetric nature of tip-sample junction.

At the lowest forces, and at low bias we observe the conduction mechanism transition to direct tunneling, as discussed more below. Furthermore, it can be observed that as the applied force increases, the curves become more symmetric in shape and exhibit natural log-like behavior at both positive and negative bias voltages indicating an ohmic-like conduction mechanism. In the log-log plot, the highest force becomes a significantly more linear in shape reinforcing the assignment as an ohmic-like behavior across the entire bias range, see Figure 5(b,c). We plot both positive and negative biases in Figure 5(b,c). The log-log data has the Fermi level set to zero with arbitrary vertical offset to set the data into a waterfall plot for ease of comparison.

Likewise, for the $Bi_2Se_3$ termination, we observe a transition in conduction mechanism as a function of force, which can be seen in our semi-log and log-log I-V spectroscopy in Figure 5(d-f). At the lowest forces, the curves show a ~4 eV band-gap like feature, which is symmetric in shape. This persists until the highest loading force, where the gap disappears and the curve exhibits log-like behavior at both positive and negative bias voltages, which indicates an ohmic-like conduction mechanism. When plotted on a log-log scale, Figure 5(f,g), the transition to an ohmic conduction is apparent as the data is no longer flat until the band edge, but the entire data set is now linear in shape across the entire bias range.

As shown on the semi-log, and log-log data in Figure 5, that at low forces the conduction mechanism is not ohmic. Plotting the data on a log-inverse scale allows additional insight into the conduction mechanism as the bias and applied force are varied. *Figure 6*(a,d) shows the plot of $\ln(|I|/V^2)$ against $1/V$ which enables t distinction of the direct tunneling (DT) and Fowler-Nordheim tunneling (FNT) regimes. The relationship between the DT and FNT current and the bias voltage can be expressed by the following theoretical equations [25–27].
g

$$\textbf{\textit{Direct Tunneling}} \; (V < V_t): I \propto V \exp\left(-\frac{2d\sqrt{2m^*\varphi}}{\hbar}\right)$$

$$\textbf{\textit{FN Tunneling}} \; (V > V_t): I \propto V^2 \exp\left(-\frac{4d\sqrt{2m^*\varphi^3}}{3\hbar eV}\right)$$

Where $\hbar$ is reduced Planck's constant, e is the electron charge, $m^*$ is the effective mass of an electron, $\varphi$ is the barrier height, and d is the thickness of the sample. The different function forms of the two regimes enable the tunneling regimes to be distinguished. When $\ln(|I|/V^2)$ versus $1/V$ is plotted, the DT should yield a natural log, while the FNT should yield a linear response.

Figure *6* shows the $\ln(|I|/V^2)$ versus $1/V$ graph as a function of applied force for both terminations. This graph exhibits a noticeable transition at a specific voltage ($V_t$), which divides the plot into two distinct regions corresponding to FN tunneling and DT, as indicated by the dashed red line in Figure *6*(a,d). This transition between the two tunneling modes displays a characteristic shape such that there is a steep linear shape at high bias voltages (FNT), followed by an inflection point that changes the shape of the curve to a natural log functional form (DT) [26,27]. This we conclude, at low bias voltages and low forces the conduction mechanism is DT, until a transition voltage $V_t$ where the mechanism changes to FNT. The asymmetry of the I-V curve is pronounced in the log-inverse plotting of the bismuthene termination (Figure *6*(a)), but since the conduction mechanism at low forces can be assigned to direct tunneling, the asymmetry can be tentatively assigned to the topological-trivial metal nature of the tip-sample junction. The general well known rectifying behavior of the tip-sample junction may also play a role.

Figure *6*(b,c,e,f) shows the graph of $\ln(|I|/V^2)$ versus $\ln(1/|V|)$, which is plotted to linearize the data in the low voltage region to further emphasize the DT nature of the conduction mechanism, and elucidate the conduction at high loading forces. As the force increases, the tip sample junction becomes increasingly more ohmic in nature, i.e. becoming closer to a constant function which is proportional to conductance. Interestingly, the observed conduction mechanism we see does not follow what would be expected of a simple semiconductor metal junction, which would have diode-like behavior resulting from Schottky and thermionic emission. C-AFM studies on $Bi_2Te_3$ observed I-V curves of similar shape as those displayed in Figure *6* at high load forces under larger biases, and the I-V curves were fit with the thermionic emission and Fowler-Nordheim tunneling model [30]. The work on $Bi_2Te_3$ shows that using the TE model, the ideality factor is indistinguishable between I-V curves acquired at step edges and at terraces step edges and terraces, however the extracted barrier height is slightly reduced at the step edges compared to the terraces. Due to the high ideality factor values, the exponential behavior, and the near symmetry of the I–V curves, they proposed that the I–V characteristics are influenced by a tunneling mechanism. The barrier heights obtained using the FNT model were found to be statistically equivalent at both the step-edges and the terraces.

In our measurements performed in the low bias and low force regime, the conduction mechanism has been shown to be direct tunneling as the force increases the domain over which the DT dominates the conduction mechanism shrinks until a transition to ohmic conduction is observed as a function of force. In the DT (low bias and force), regime the differential conductance (dI/dV) spectroscopy reflects the sample local density of states (LDOS), and so is a probe of the Dirac cone of the bismuthene termination. The distinction between the DT and Ohmic regime can be observed in *Figure 7*. In Figure *7*(a,d) plots dI/dV against V as a function of applied force across a ± 2 V range, where the conduction mechanism transitions from DT to Ohmic, where the dI/dV signal is not strictly proportional to the LDOS. In contrast, Figure *7*(b,e) displays the dI/dV acquired over a smaller voltage region, where the DT mechanism dominates the conduction, and the LDOS can be examined around the Fermi level. The shape of the dI/dV spectra show a force dependent, which is consistent with the transition from the DT to Ohmic conduction regimes. At the low forces and biases, the spectra are expected to reflect dI/dV spectra taken from the previous STM study [8]. For dI/dV of the bismuthene terminated surface, the voltage range that the dI/dV spectroscopy is linear in shape, decreases as the applied force increases. Likewise, the Fermi-level approaches zero volts as the force increases. We anticipate a Dirac-cone around the Fermi-level for the bismuthene terminated surface as found previous in STM research [8]. For the lightest forces applied 15 nN and 46 nN, in Figure *7*(b) we observed the linear dI/dV across ±1 V with the Fermi-level ~0.4 V. From 77 nN to 154 nN, we observe the Dirac cone from ±0.5 V, with the Fermi level shifting from ~0.2 V at 77 nN to ~0.1 V at 154 nN. Compared to the previous STS which observed the Fermi-level at zero volts, and the Dirac-cone covering a range from -0.4 V to +0.2 V, our results qualitatively agree well; with mid-force spectra in excellent agreement. Conversely, at the highest force, 230 nN, the Dirac cone is no longer visible, with a parabolic-shaped dI/dV. We know from the above analysis at this force the conduction mechanism is no longer direct tunneling at low bias, and switches to an ohmic conductions.

Differential conductance spectra presented in Figure *7*(c,d) show a force-dependent evolution. At low forces, the dI/dV curve exhibits a suppressed conductance around the Fermi level. As force increases, the gap decreases, until at the highest applied force where the gap vanishes entirely, which also corresponds to the transition to a more ohmic contact regime. The dI/dV spectroscopy is featureless within the gap, in contrast to bismuthene, which has a clear linear Dirac dispersion around the Fermi level. At the lowest applied forces (15 nN and 46 nN), the spectra exhibit a broad gap that is symmetric with respect to the Fermi level. Literature reports of measurements performed on $Bi_2Se_3$ stoichiometric films report a small gap of ~0.3 eV [41,42], however *Figure 7*(c,d) shows a significantly larger gap-like feature of several eV, which is tentatively explained by tip-induced band bending, which is known to exaggerate electronic gaps in semiconductors.

In order to estimate the bandgap from the I-V spectroscopy, the data fit with the thermionic emission model discussed above, resulting in an average gap of ~0.41 eV, which is consistent of literature reports [41,42]. In contrast to prior scanning tunneling spectroscopy studies of the $Bi_2Se_3$ terminated surface, which showed a parabolic dispersion in the dI/dV [8], the dI/dV plotted in Figure *7*(c,d) shows a gapped feature. The prior STM studies were performed under UHV conditions maintain a pristine surface, where the measurements reported here were reported under ambient conditions where the role of oxygen and adsorbed water on the $Bi_2Se_3$ is relatively unknown, but may influence the LDOS. However, as the applied force increases (at 200 nN and 230 nN), the shape of the dI/dV spectra gradually transitions from having a large gap to showing a parabolic shape with a small gap or no gap at all. As was the case with the STS study, there is not strong spectroscopic signature of the topological state for the $Bi_2Se_3$ termination which is located within the valence band (around -0.6 eV) of the Bi$_2$-Bi$_2$Se$_3$ band structure[8]. Which does not indicate the absence of the topological state, but that rather the signature maybe buried with signal from the valance band states.

Overall, the influence of force-dependent tip-sample interaction underscores the importance of maintaining weak contact conditions, to remain in the DT regime, when probing the intrinsic density of states near the Fermi level in C-AFM studies. By doing so, one can access the LDOS even under ambient conditions and directly probe the topological surface states without the need for more experimentally demanding UHV conditions. Furthermore, this work is an important proof of principle study for future device processing schemes under typical photolithographic recipes, as it has been shown that the topological nature of Bi$_2$-Bi$_2$Se$_3$ system to be robust to ambient conditions.

**Bi$_2$-Bi$_2$Se$_3$ Edge State Imaging:**

C-AFM was performed on an atomically smooth nanopyramid of Bi$_2$-Bi$_2$Se$_3$ and observed enhancement in conductivity at the step edges was observed, which is similar to what has been observed by C-AFM in Bi$_2$Te$_3$ [30]. Figure *8* shows a topographic and current image of a hexagonally shaped nanopyramid with a pair of screw dislocation at its apex. The current image shows a clear enhancement in signal at the perimeter of each terrace, which is visible as the bright yellow-orange contrast in Figure *8*(b). This edge state is spatially robust, and spirals outward over dozens of terrace steps.

In order to probe the spatial and energy variations of the edge states, current imaging was performed under a variety of force and voltage setpoints. *figure 9* (a) shows the topographic image of a terrace staircase, while *figure 9*(b) displays the simultaneously acquired current map, where the data was acquired in the high force (154 nN) and low bias (+0.1 V) condition. The current map shows enhanced current that is localized at the step edges and is visible as the bright yellow-orange contrast. Little to no current is measured within the terraces despite the high force setpoint. Interestingly, with this load force an effective interaction area between the tip and sample of 48 nm$^2$ is expected which is much larger than the width of the conducting edge state yet the current is still highly localized at the terrace/step edges. Alternatively, under low force (7 nN) but high bias (+2 V) conditions the edge state remains localized on at the terrace perimeters, see *Figure 10*(a). Under these bias and force conditions the expected interaction area is 11 nm$^2$ which is commensurate to the width as the edge state. Although it is worth noting, the effective electrical interaction (see supplement), does not take into account effects like the water meniscus or real tip geometry, all of which influence the collection area. While generally speaking, a high force enables a larger effective collection area.

Step height analysis of the staircase shown in Figure *9*(a) reveals the terminations to be a combination of bismuthene termination and Bi$_2$Se$_3$ terminations [8,10]. Figure *9*(c) shows a series of linecuts used to determine the terrace terminations, with some lower terraces partially terminated by bismuthene, which is indicated by the orange and red lines showing of a step height of ~0.3 nm, consistent with the Bi$_2$ step height. Moving up the staircase, the termination switches to a $Bi_2Se_3$ termination for the topmost terraces, see the blue line in Figure *9*(c). Comparatively, in C-AFM studies of $Bi_2Te_3$ there is only one termination type, and they observe enhanced conduction at the edges of the terraces, similarly in bismuthene research by STM on $SnSe_2$ enhanced conductivity is observed as an edge state of the 2D sheet[45]. Our work reflects both cases and shows enhanced current contrast along the edges of the either termination in bismuthene or $Bi_2Se_3$.

As shown by the I-V spectroscopy presented above, at low forces minimal current flows through the bismuthene terminated terrace, while at low biases, minimal current flows through the $Bi_2Se_3$ terminated terrace. As force or voltage are gradually increased, eventually both terminations will become conductive. As a result, it can be expected that there exists a combination of force and bias setpoints under which the center of the terraces, regardless of termination, will become conductive. The current imaging was investigated as a function of the force×voltage product, and as the product grows in magnitude, the edge state increases in apparent width until the entire terrace is conductive.

Figure *10* shows an array of current images with corresponding linecuts of multiple terraces with increasing product of force×voltage. Initially, the current is localized to only the perimeter of terrace with a width of approximately 25 nm, see Figure *10*(a,b). The width of this edge state may be overestimated by the finite size of the AFM tip, which has a nominal radius of 25 nm. The width of edge state begins to increase at a force×voltage product of ~70 nN·V, and the apparent width of the edge stage grows until at a force×voltage product of ~185 nN·V, the entire terrace appears to be conducive. Figure *10*(h) plots the edge state widths, as extracted from the linecuts in *Figure 10*(a-g), as a function of the force×voltage product. It is worth noting that the average terrace width size is ~70 nm, and the trend in the force×voltage product tends towards this value, as indicated by the dashed line in Figure *10*(h).

If we compare our results to other studies in bismuthine, we find results which indicate the edge states on bismuthine are topological. However, the manifestations of the topological nature of bismuthine is strongly influenced by the choice of substrate. For monolayer bismuthene on $SnSe_2$ researchers demonstrated the realization of a 2D Weyl semimetal. Through spin and angle-resolved photoemission and STS, the researchers observed both the linear Weyl dispersion and topologically protected Fermi string edge states [45].

For the case on bismuthine on Ag(111) researchers attributed the enhancement of conductivity at terrace edges to the strong SOC of bismuth and its orbital hybridization with the Ag(111) substrate. It induces an orbital-filtering effect in the bismuthene, resulting in a large topological gap of around 1 eV. STS measurements revealed Dirac cone-like features in the LDOS, which is consistent with the presence of topologically protected edge states [43].

In a study on bismuthene grown on a SiC(0001) substrate, researchers successfully synthesized it as a monolayer honeycomb lattice, forming a highly ordered 2-D material. STS measurements revealed a large electronic bandgap of approximately 0.8 eV, supported by ARPES measurements and DFT calculations, while confirming the presence of conductive edge states. The electronic structure is dominated by strong on-site SOC, along with hybridization with the substrate [44].

In order to more thoroughly investigate the electronic nature of the edge-states, force dependent I-V spectroscopy on both terminations was performed. Figure *11*(a) shows an atomic staircase with a corresponding line cut shown in Figure *11*(b), where the black X marks the location where the force dependent I-V spectroscopy, plotted in *Figure 11*(c), was performed. For bismuthene or $Bi_2Se_3$ terminations, as the applied force increased, the current also increases—however for the edge-state the opposite trend holds, where increased force begins to open an electronic gap and decrease the conductance. The measurements were conducted in the low force regime, as with sufficiently high forces, the effective interaction area between the tip and sample increases, and the edge states can no longer be probed independently of the terrace. The data were plotted on semi-log and log-log axes in order to identify the function form and resulting conduction mechanism, see *Figure 12*. In the semi-log plot the lowest forces resulted in a sharp V-shape, however, as the force increases a gap-like feature appears. We suspect the emergence of the apparent gap as the load force increases is not due to strain affecting the edge-state, as we clearly observe edges states at high loading forces and low bias (Figure *10*) but rather, the effective area of the current collection has grown sufficiently large that the signal is dominated by contributions from within $Bi_2Se_3$ terrace.

When the $\ln(|I|/V^2)$ is plotted against $1/V$, there is a visible transition between the DT and FNT regimes The DT regime is not as purely a natural log in functional form, and thelinear range for FNT regime is relatively small. However, the DT regime is more easily visible when $\ln(|I|/V^2)$ is plotted against $\ln(1/|V|)$ plotting, where a clear linear response is visible for all three force setpoints. The differential conductance is plotted in the DT regime to show the LDOS of the edge state, see Figure *13*. The LDOS is featureless, beyond the existence of states around the Fermi level. The effective conduction area of the C-AFM tip makes interpretation of the edge state difficult, as only at the lowest forces is the current collection confined sufficiently to the terrace edge.

The terrace edges may be decorated with water or oxygen which could influence the local electronic environment and perturb, or even induce these edge states that are observed. To work towards disentangling this question, we turn to cryogenic UHV STM measurements were to supplement the C-AFM measurements that have already been presented. dI/dV maps were acquired simultaneously with the topographic images, and surprisingly edge states on pristine bismuthene and $Bi_2Se_3$ were not observed. *Figure 14(a)* shows an STM image with atomic staircase that has both terminations. A linecut across the staircase is shown in *Figure 14(b)* with step heights labeled to determine termination. Corresponding dI/dV imaging under a systematic series of bias conditions of the same location as in *Figure 14(a)* is shown in *Figure 14(c-i)*. Linecuts are across the dI/dV images are shown as insets beneath each image, *Figure 14(c-i)* crossing both termination and do not show any indication of edge-states for either termination. This is in contrast to previous bismuthene STM studies demonstrating edge-states, and observing a Dirac cone within the edge-state electronic structure[43]. From this comparison of C-AFM and

STM, this points to ambient environment playing role in stabilizing an edge state on the $Bi_2$-$Bi_2Se_3$ system. At the same time, it is also clear the ambient conditions do not affect the topological nature of the Bismuthene within the terrace as the Dirac cone is still clearly observed.

To more robustly probe this phenomena additional STM measurements should be performed, where a pristine surface is produced, and subsequently in-situ dosing of oxygen is allowed onto the sample to understand the incorporation.

**Conclusions**

Our study demonstrates that $Bi_2$–$Bi_2Se_3$ superlattices preserve their topologically protected states under ambient conditions, with force-voltage-dependent conducting AFM providing valuable insights into the electronic transport properties of these materials in complex environments revealing a crossover from direct tunneling, to Fowler-Nordheim, and up to ohmic-like transport. The observation of Dirac cone features, and robust edge conduction, underscores the resilience of both surface and edge states to environmental perturbations. These results establish $Bi_2$–$Bi_2Se_3$ as a promising platform for ambient-compatible quantum devices and highlight the utility of functional scanning probe methods for probing fragile quantum states in realistic conditions.

**ACKNOWLEDGMENTS**
This material is based up work supported by the National Science Foundation under Grant No. 2328747. STM experiments were conducted at the NSF NeXUS facility at The Ohio State University. Financial support was provided by NSF Award 2410901.

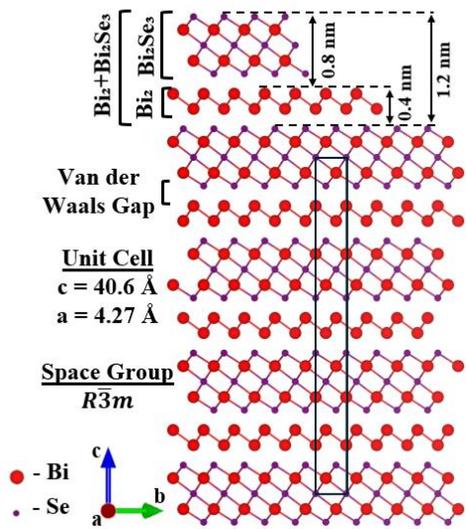

Figure 1 Crystal structure of $Bi_2$-$Bi_2Se_3$. The $Bi_2$-$Bi_2Se_3$ superlattice is composed of a 2D $Bi_2$ layer and a quintuple-layer (QL) slab of $Bi_2Se_3$. The layers are separated by van der Waals gaps.

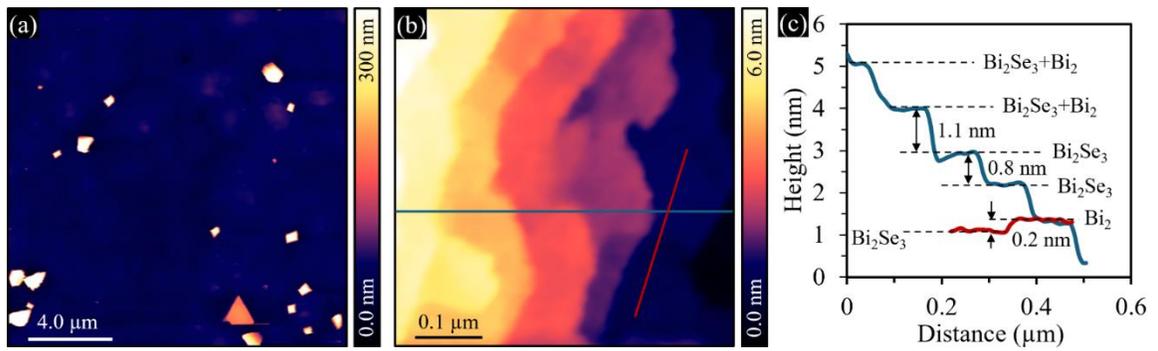

Figure 2 AFM images and linecuts. (a) Isolated crystal forms embedded within a background of atomically smooth layers. (b) Zoomed-in view of a 0.5 μm × 0.5 μm region showing the atomic staircase with 001-layer stacking. Line cuts are taken across (blue) and along (red) the staircase. (c) Characteristic step height derived from the line cuts in (b).

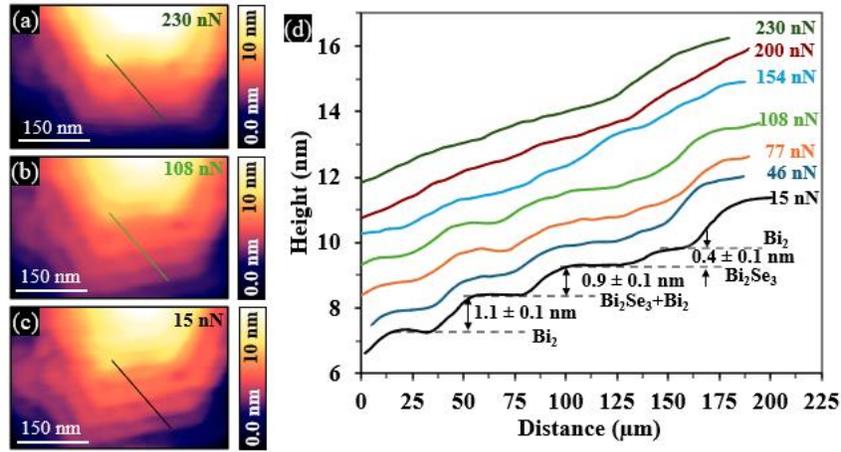

Figure 3 AFM imaging and linecuts. (a-c) AFM topographic images under different force conditions shows a blurring of topographic information. Force conditions are colored and labeled. (d) shows linecuts at various force conditions increasing from 15 nN to 230 nN. Lines shown in (a-c) indicate the position of linecuts.

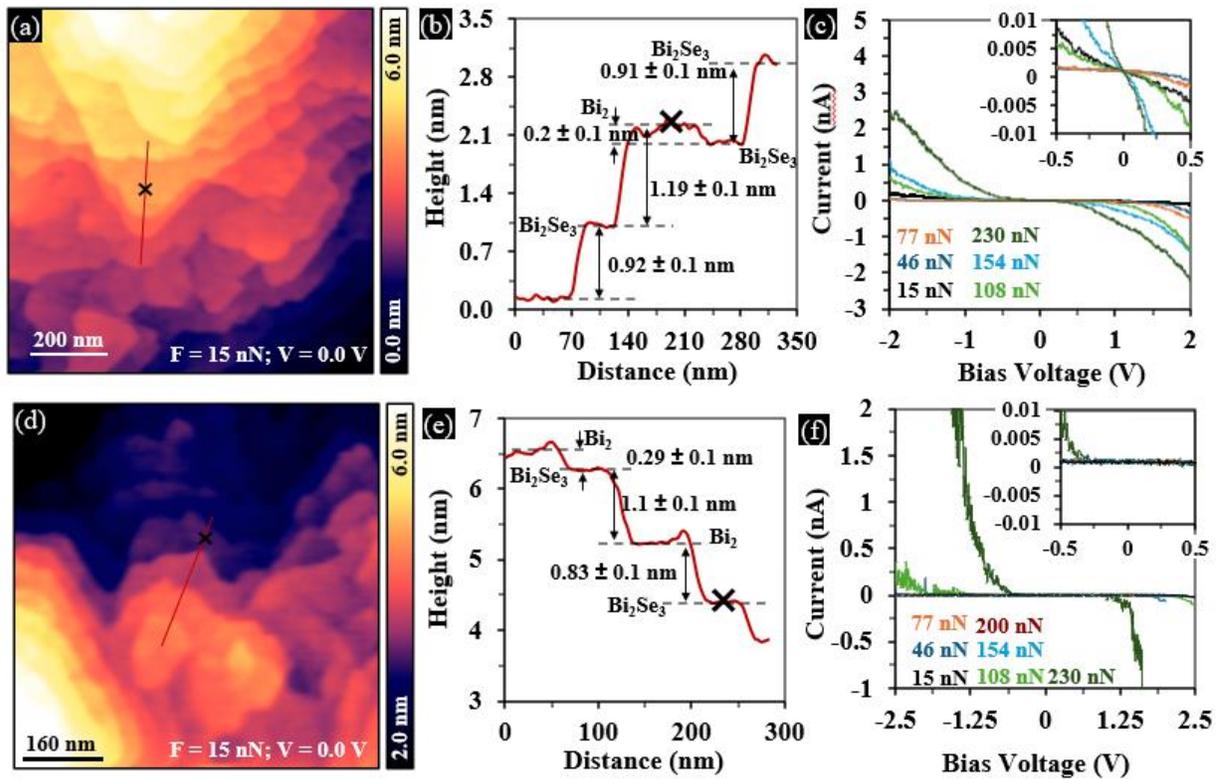

Figure 4 C-AFM and IV spectroscopy of the Bi$_2$-Bi$_2$Se$_3$ thin film. (a,d) C-AFM topographic image showing atomically smooth steps in a 1 μm × 1 μm region, with a line cut across the staircase indicated by the red line. The location of the IV spectroscopy is marked by the black X; Bi$_2$ termination in (a), and Bi$_2$Se$_3$ termination (d). (b,e) linecut showing characteristic step heights (a,d) to determine the termination. A 0.2 nm step height corresponds to a single Bi$_2$ atomic layer, while a QL step of Bi$_2$Se$_3$ measures approximately 0.8 nm. I-V point spectroscopy is shown for both termination; (c) Bi$_2$ termination and (d) Bi$_2$Se$_3$ termination.

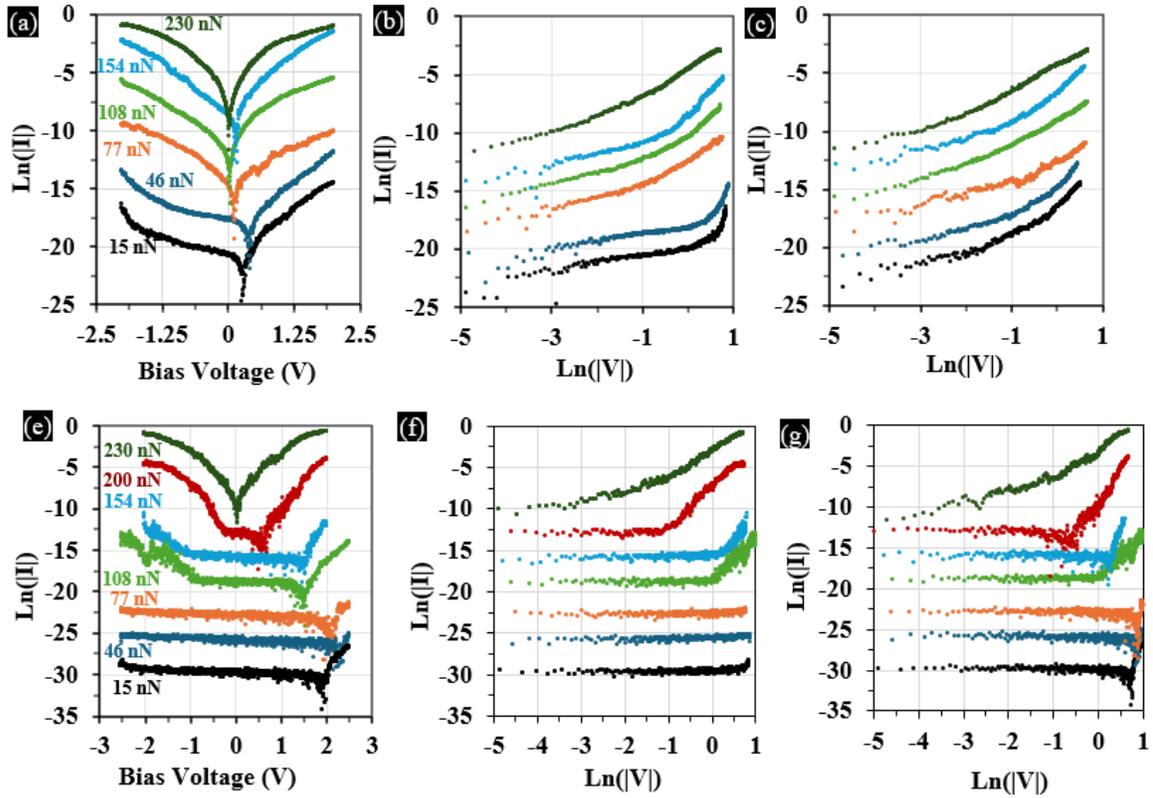

Figure 5 I-V point spectroscopy data as a function of force (15 nN to 230 nN) graphed in semi-log and log-log plotting. (a,e) semi-log data corresponding to the bismuthene (a), and $Bi_2Se_3$ (e) terminations as indicated in Figure 4. (b,c,f,g) Log-log data curves as a function of applied force between are displayed for the positive bias voltages in (a,f) and the negative bias voltages in (c,g) for the bismuthene (a), and $Bi_2Se_3$ (e) terminations. Curves are arbitrarily offset vertically for comparisons.

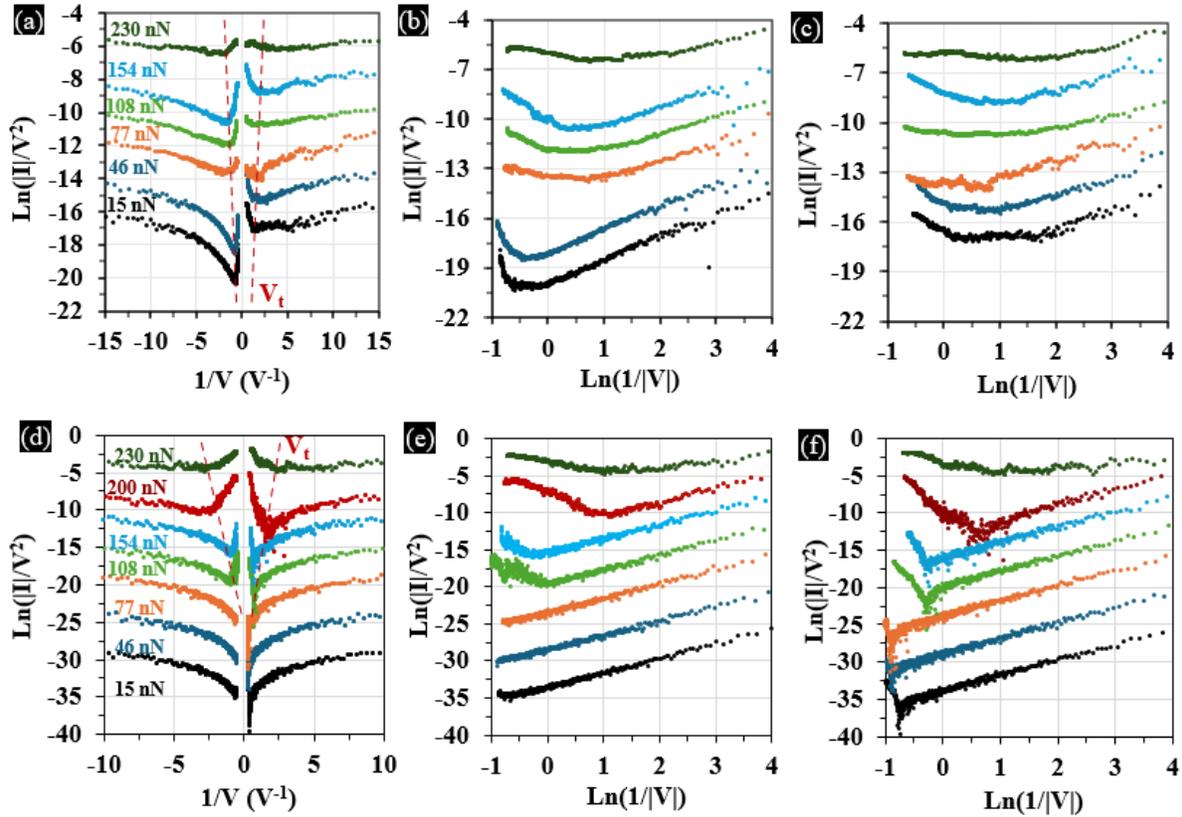

Figure 6 Additional log-inverse and log-log plotting for IV-spectroscopy for both $Bi_2$ and $Bi_2Se_3$ terminations. (a,d) $Ln(I/V^2)$ vs. $1/V$ curves as a function of applied force between 15 nN and 230 nN show the transition between FN tunneling and direct tunneling for Bi2 (a) and Bi2Se3 (d). At low forces (15 nN and 46 nN) and in the high-bias voltage region, the linear relationship confirms FN tunneling. Fermi-level was shifted to zero prior to plotting $Ln(I/V^2)$ vs. $1/V$ such that the transitions from tunneling to FNT are lined up for all curves for display purposes only. $Ln(|I|/V^2)$ vs. $Ln(1/|V|)$ curves as a function of applied force between 15 nN to 230 nN in the positive bias voltage region are shown in (b,e) and in the negative bias voltage region in (c,f). In (b,e) these curves show a linear relationship at low forces (15 nN and 46 nN) and in the low-bias voltage region, confirming direct tunneling. Where at the highest force (230 nN) the curves become more most constant indicating a transition towards a ohmic-like contact. Curves are arbitrarily offset vertically for comparisons.

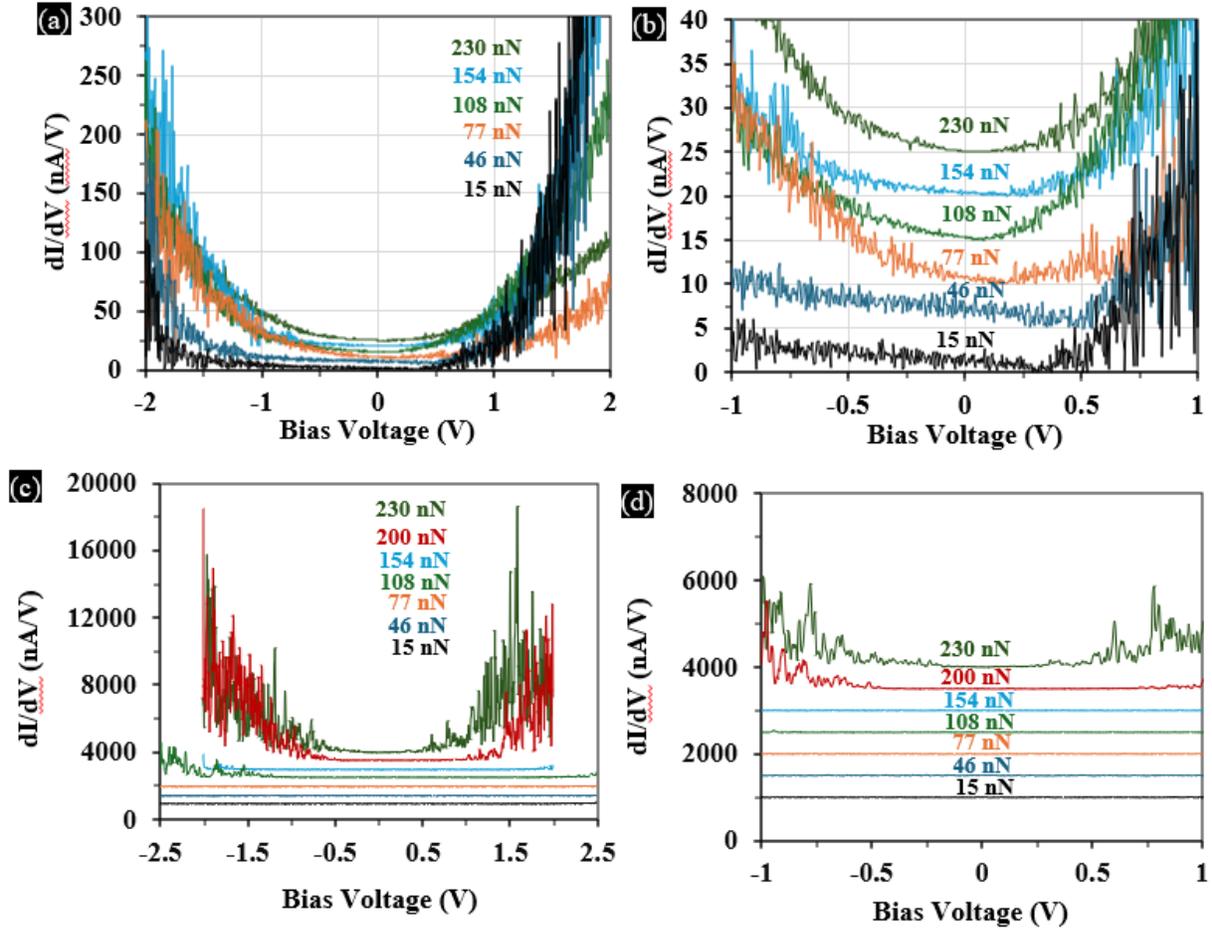

Figure 7 dI/dV vs. V curves as a function of applied force. Each curve is scaled by an arbitrary factor (15 nN: 30, 46 nN: 50, 77 nN: 15, 108 nN: 4, 154 nN: 8, and 230 nN: 1) and vertically shifted by adding an arbitrary offset for comparisons. (a,c) dI/dV vs. V for the bismuthene (a) and Bi2Se3 (c) terminations. (b,d) Zoomed-in view of plot (a,c) to show the states around the Fermi-level. In bismuthene terminations the Dirac cone is observed at the lowest forces, 15 nN, 46 nN, and 77 nN. Whereas the $Bi_2Se_3$ termination shows a wide band-gap until the highest forces. Curves are arbitrarily offset vertically for comparisons.

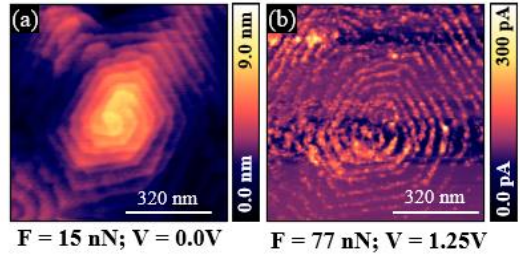

Figure 8 topographic (a) and simultaneous current imaging (b) of a hexagonal nano-pyramid with a screw dislocation at it apex. Localized current is observed as bright (orange-yellow) contrast localized at the edges of the terraces.

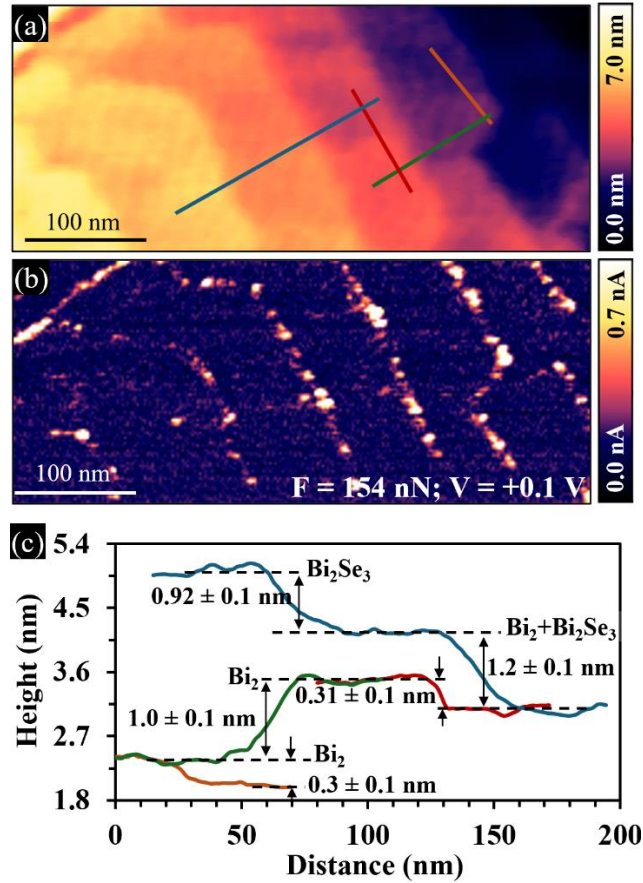

Figure 9 C-AFM Images and linecuts. (a) AFM topographic image of $Bi_2$-$Bi_2Se_3$ terraces with both terminations. Linecuts (c) are taken along the colored lines and displaced in (a). (b) Corresponding C-AFM image showing edge current with set conditions: +0.1 V bias, 154 nN applied force. Edge state current can be seen as the bright, orange-colored lines within the dark purple background. (c) linecuts taken from (a) corresponding to a series of step heights along and within the terraces. Terminations and step heights are labeled within the graph.

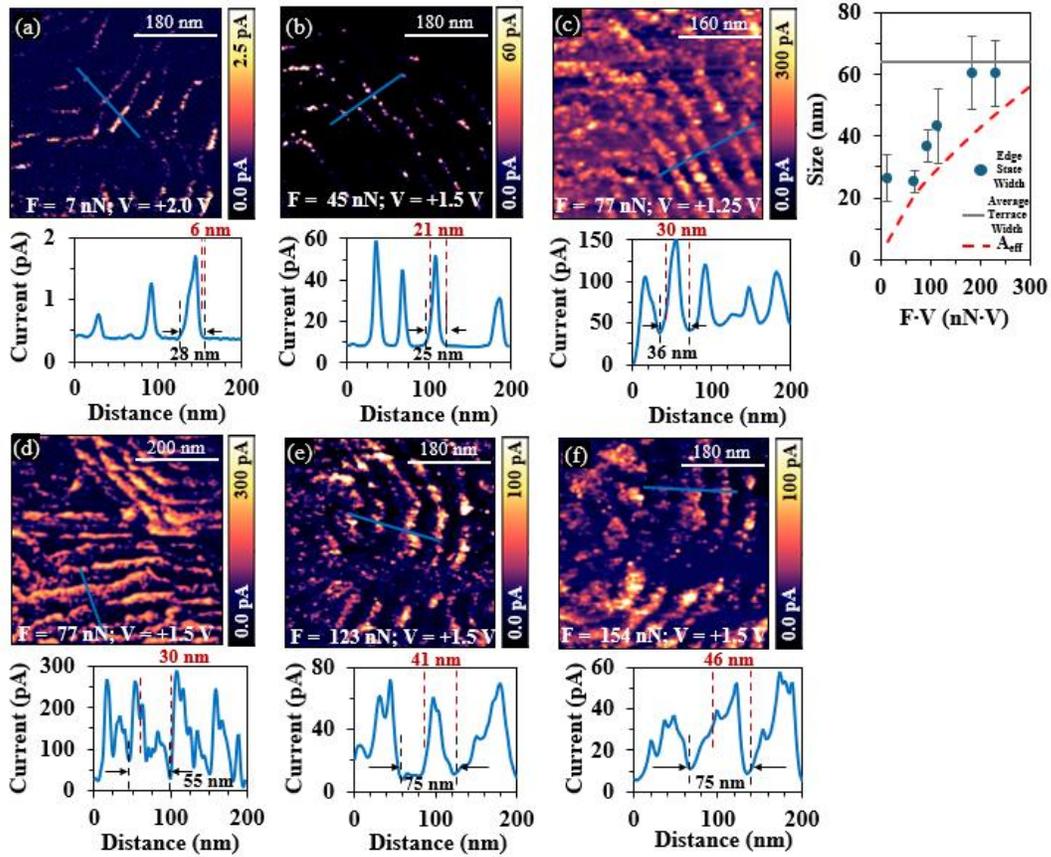

Figure 10 current images and linecuts under a systematic combination of force and bias voltage to measure the change in the edge state. Blue lines in (a-f) are position of linecuts, with the corresponding linecut shown below the image. Red dashed lines indicate the size of the effective conducting area, where the black dashed line indicates the width the conducting area. (h) is a graph of the edge state width versus the product of force and voltage. We observe the edge state width initially doesn't change until an onset where it progressed linearly until it levels off at the average terrace width size.

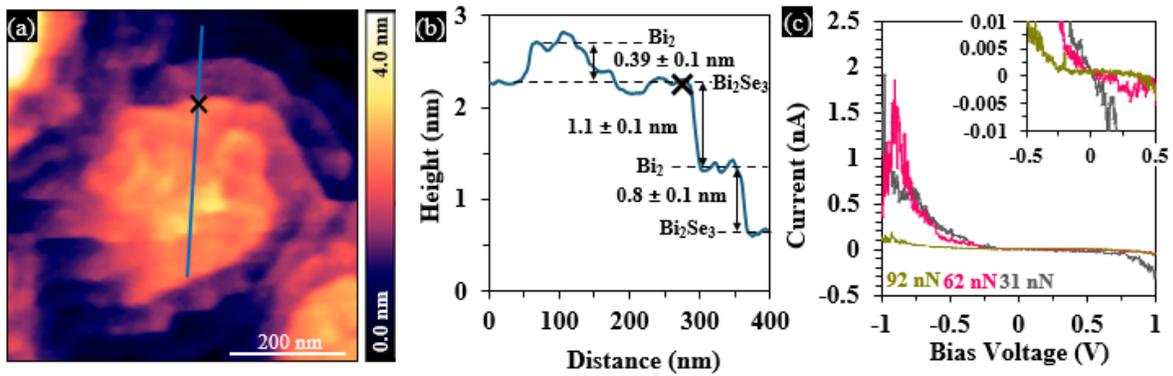

Figure 11 C-AFM and IV spectroscopy of the $Bi_2$-$Bi_2Se_3$ thin film. (a) topographic image showing atomically smooth steps in a 1 μm × 1 μm region, with a linecut across the staircase indicated by the blue line. The location of the IV spectroscopy is marked by the black X. The linecut is displayed in (b) with the termination assignments labeled within the graph. IV-point spectroscopy is shown in (c) with a inset showing a close-up around zero bias.

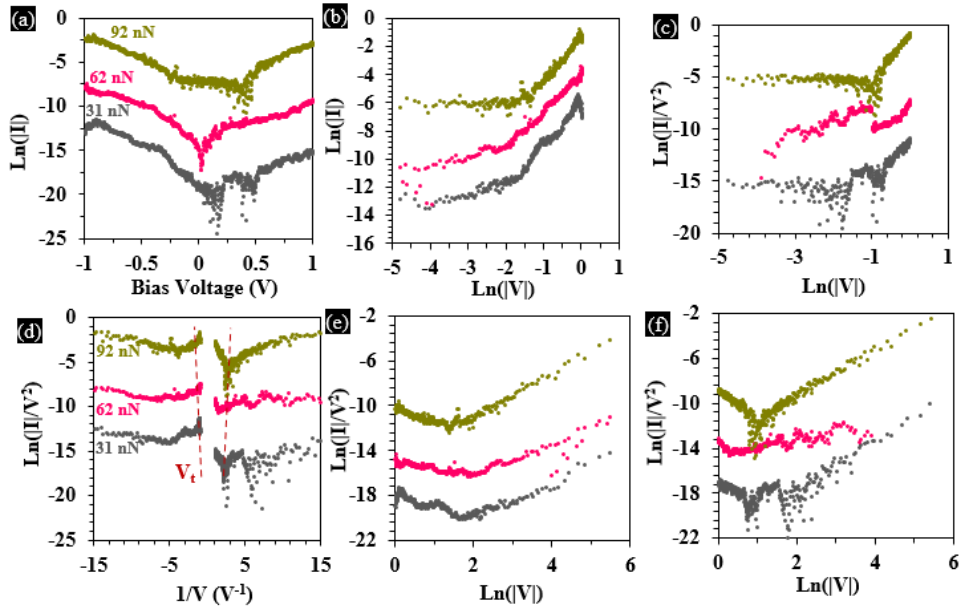

Figure 12 semi-log, log-log and log-inverse plots of the IV-point spectroscopy data of the edge state. Position of spectroscpy indicated in Figure 11. (a) semi log data of the IV curve as a function of force ranging from (31 nN to 92 nN). At the lowest forces DT characteristic dominate the functional shape, where at 92 nN a gap emerges in the data. (b,c) log-log plots of IV curves indicating DT at low bias and forces below 92 nN. (d) log-inverse plots show potential emergence of an FNT characteristic at the highest loading force. Curves are arbitrarily offset vertically for comparisons.

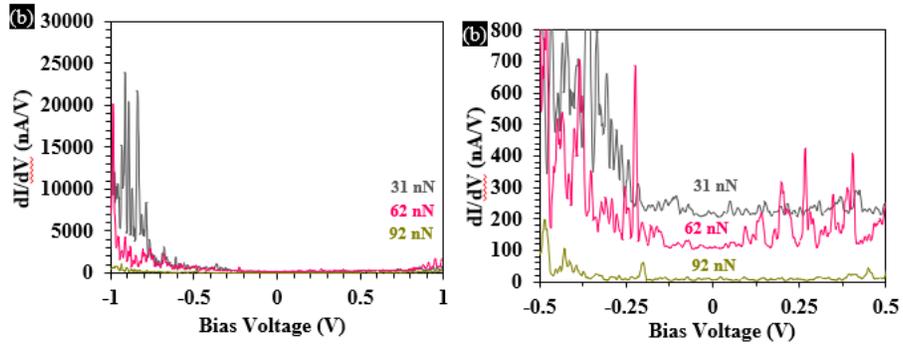

Figure 13 dI/dV vs. V plots of the edge state as a function of force shows a gap emerging as the force increases to 92 nN. Where at lowest forces states exist around the Fermi-level. Curves are arbitrarily offset vertically for comparisons.

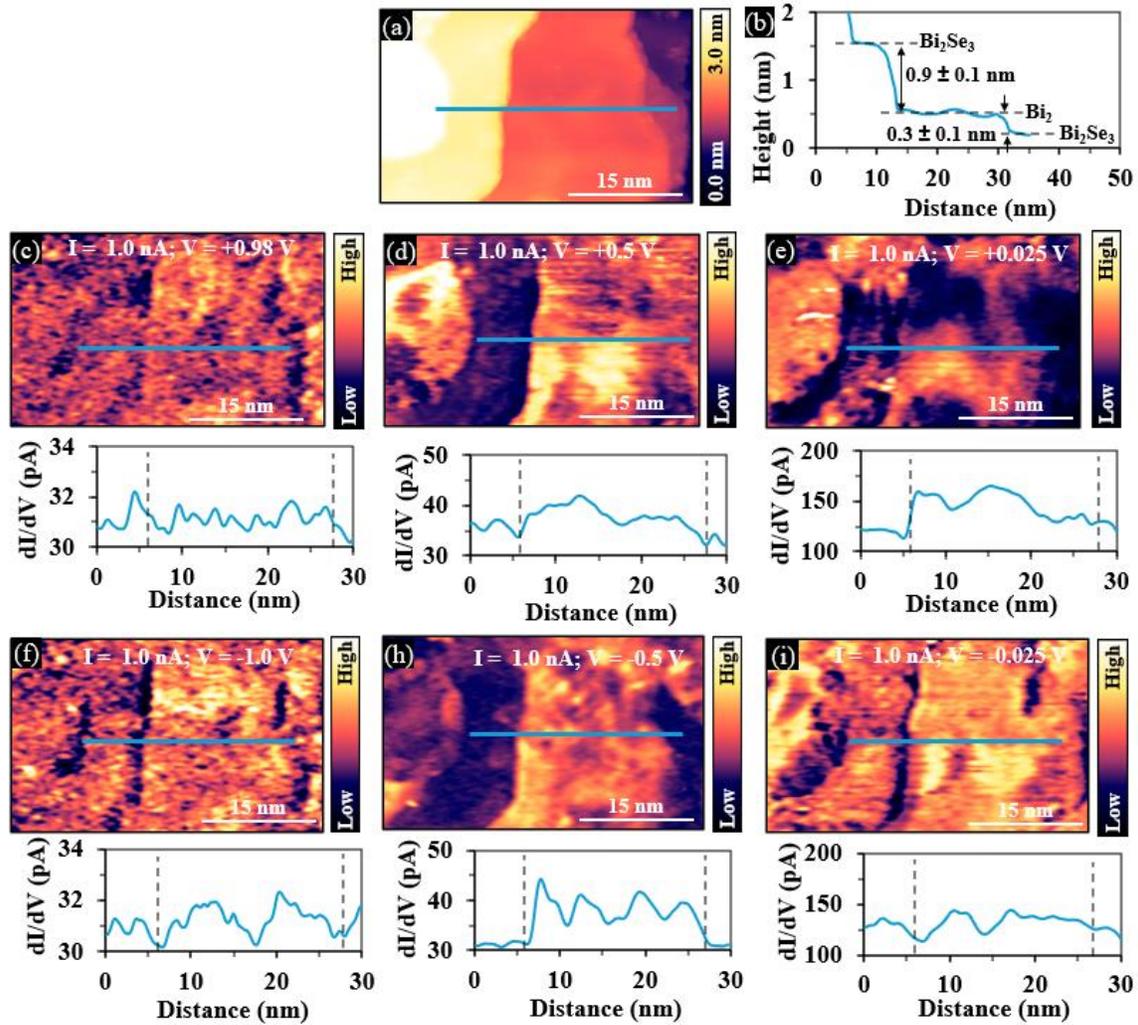

Figure 14 STM topographic and dI/dV imaging. (a) topographic image of an atomic staircase with the blue line indicating the position of a linecut shown in (b). (b) linecut from (a) showing a $Bi_2Se_3$ and $Bi_2$ termination. (c-i) dI/dV imaging under systematic voltage variation (±1 V). Electronic contrast emerges between terminations for specific biases, but no indication of a sharp localized edge emerges in the imaging. Linecuts taking across the dI/dV images spanning both terminations is shown below each dI/dV image.

# SUPPLEMENTAL: Robust Topological Conduction in Bi2–Bi2Se3 Superlattices at Ambient Conditions

Lakshan Don Manuwelge Don[1], Md. Sakauat Hasan Sakib[1], Gracie Pillow[1], Sara McGinnis[1], Seth Shields[2], Joseph P. Corbett[1]
1. Miami University, Department of Physics, Oxford, OH 45056
2. NSF NeXUS Facility, The Ohio State University, Columbus, OH, 43210

**X-ray Diffraction Data:**

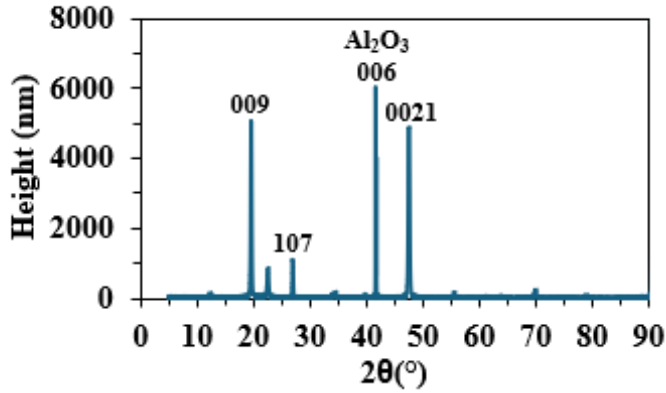

Figure S 1 X-ray diffraction data from a couple symmetric scan of Bi2-Bi2Se3 film showing 001 oriented films.

**Effective Electrical Interaction Area:**

In C-AFM the effective interaction area of the probe depends on the mechanical properties of both the tip and the sample, and primarily on the applied force, as given by the formula [25,30]:

$$A_{eff} = \pi \left( FR_{tip} \frac{3}{4} \left( \frac{1 - V_{tip}^2}{E_{tip}} + \frac{1 - V_{sample}^2}{E_{sample}} \right) \right)^{\frac{2}{3}}$$

Where F is the contact force, $R_{tip}$ is the tip radius, and $E_i$ and $V_i$ are Young's modulus and Poisson's ratio of tip and sample material. For our lightest forces for Pt-coated tips, an effective tip interaction area of ~10 nm is computed, whereas at our largest force and effect tip interaction area dramatically increases to about ~60 nm, see Table 1, the change in effective area as a function of force, although these computations are at best estimates, the actual effective interaction area is difficult to ascertain, as it depends on the specific geometry of the tip apex and the specific gaseous environment of the tip. Under ambient conditions, it is assumed a small meniscus of water is formed which in general increases the effective area.

Table 1. The change in effective area of the Pt-coated Si probe as a function of applied force.

| Applied Force (nN) | 15 | 46 | 77 | 108 | 154 | 200 | 230 |
|---|---|---|---|---|---|---|---|
| Effective Area (nm²) | 10 | 22 | 30 | 38 | 48 | 58 | 63 |

**Thermionic Emission model:**

While we do not expect a thermionic emission given the above analysis, we nevertheless, we fit a thermionic emission model of metallic-semiconductor junction to the high bias regime of the semi-log data to exact out barrier heights and compare with C-AFM Bi$_2$Te$_3$ results. The thermionic emission equation and the diode equation are given by [32]:

$$I_0 = A_{eff} A^{**} T^2 exp\left(-\frac{q\Phi_B}{k_B T}\right)$$

$$I = I_0 exp\left(-\frac{qV_{bias}}{\eta k_B T} - 1\right)$$

Where $I$ is the current, $I_0$ is saturation current, $q$ is the electron charge, $V_{bias}$ is the bias voltage, $\eta$ is the ideality factor, $k_B$ is the Boltzmann constant, $T$ is the temperature, $A^{**}$ is the Richardson Constant, $\Phi_B$ is the barrier height. Under these conduction mechanisms one expects the IV curve to be linear in a semi-log plot. While we do not strictly observe this, we can make estimates of the tip-sample barrier height to compare with Bi$_2$Te$_3$ C-AFM barrier heights. At high biases, the ln(I) vs. V plots were fitted to a linear relationship to extract fitting parameters for determining the barrier height and ideality factor assuming this was a thermionic emission. The calculated values for each applied force at both polarities are shown in supplemental Table Y1. The Richardson constant was assumed to be $3 \times 10^6\ Am^{-2}K^{-2}$ [33], and the sample parameters (Poisson's ratio: 0.27 and Young's modulus: $5.86 \times 10^{10}\ Nm^{-2}$) [34] were assumed to be the reasonably similar as those of Bi$_2$Se$_3$, since these parameters do not exist in the literature for Bi4Se3. Tip parameters were taken to be those of Pt (Poisson's ratio: 0.36 and Young's modulus: $1.7 \times 10^{11}\ Nm^{-2}$) [25]. The barrier heights obtained from the TE model for bismuthene termination show a decreasing barrier height as a function of applied force ranging 0.21 eV down to 0.10 eV. Where as the barrier heights for the Bi2Se3 termination ranged from ~0.77 eV at low force to ~0.17 eV at high force, slightly higher than those of Bi-terminated regions, consistent with the semiconducting character of Bi2Se3 [40]. Previous research on Bi$_2$Te$_3$ has reported two separate barrier heights, one at the center of a Bi$_2$Te$_3$ terrace with 0.351 eV and step-edge barrier height of 0.299 eV, which is in a similar ballpark. Since the bismuthene termination is metallic we expect a small barrier height to exist and to decrease with applied force as the tip makes closer contact, although it is difficult to rule out atmospheric interference that may artificially increase the barrier height. Futhermore, we do not know the effect the oxygen in the atmosphere plays on the local compositions of these materials, while from this data, and other works in Bi$_2$Se$_3$ that oxygen plays a minimal role in altering the materials behavior. Although it should be noted there is some debate in the literature regarding the effects of oxygen from the atmosphere on Bi$_2$Se$_3$ [35–38].